# Log-Periodic Precursors to Volcanic Eruptions: Evidence from 34 Events

Qinghua Lei[1,*], Didier Sornette[2]

[1]*Department of Earth Sciences, Uppsala University, Uppsala, Sweden*

[2]*Institute of Risk Analysis, Prediction and Management, Academy for Advanced Interdisciplinary Studies, Southern University of Science and Technology, Shenzhen, China*

**Abstract** Forecasting volcanic eruptions remains a formidable challenge due to the inherent complexity and variability of volcanic processes. A key source of uncertainty arises from the sporadic nature of volcanic unrest, which is often characterised by intermittent phases of quiescent deceleration and sudden acceleration, rather than a consistent, predictable progression towards eruption. This seemingly erratic pattern complicates volcano forecasting as it challenges conventional time-to-failure models that often assume a simple smooth power law acceleration. We propose a log-periodic power law singularity model, which effectively captures the intermittent and non-monotonic rupture dynamics characteristic of reawakening volcanoes at the site scale. Mathematically, generalising the power law exponent by extending it from real to complex numbers, this model captures the partial break of continuous scale invariance to discrete scale invariance that is inherent to the intermittent dynamics of damage and rupture processes in heterogeneous crustal systems. By performing parametric and nonparametric tests on a large dataset of 34 historical eruptions worldwide, we present empirical evidence and theoretical arguments demonstrating the statistical significance of log-periodic oscillations decorating power law finite-time singularities during pre-eruptive volcanic unrest. Log-periodicity in volcanoes may originate from various mechanisms, including diffusion-dominated magma flow, magma-driven propagation of subparallel dykes, interaction between stress drop and stress corrosion, and/or interplay of inertia, damage, and healing within volcanic systems. Our results have important implications for volcano forecasting, because understanding and characterising log-periodicity could turn the intermittency of volcanic activity from a challenge into a valuable asset for improving predictions.

**Keywords** Volcanic eruption; Catastrophic failure; Power law; Discrete scale invariance; Log-periodicity; Prediction

---

* Corresponding author: qinghua.lei@geo.uu.se

## 1. Introduction

Volcanic eruptions pose significant threats to life, infrastructure, and ecosystems, with impacts ranging from local catastrophic devastation to global climate disruption. Forecasting volcanic eruptions is therefore crucial and a fundamental objective of volcanology (Sparks, 2003; Acocella et al., 2023). Achieving reliable volcano forecasting is essential for enabling civil authorities to effectively inform the public about potential eruptions and their timing, ensuring timely preparation, facilitating timely evacuations and the implementation of other safety measures. Over the past decades, great efforts have been dedicated to developing and deploying high-precision monitoring technologies to observe active volcanoes (White and McCausland, 2016; Sparks et al., 2012; Poland et al., 2020), aiming to detect and characterise potential precursors of impending eruptive events as well as anticipate their subsequent evolution.

Numerous approaches, ranging from physics-based models to data-driven empirical techniques, have been developed to forecast imminent volcanic eruptions (Acocella et al., 2023; Blake and Cortés, 2018; Robertson and Kilburn, 2016; Sparks, 2003; Voight, 1989, 1988). Among these, the Failure Forecast Method (FFM), which integrates a power-law time-to-failure model (Voight, 1989) with an inverse rate analysis technique (Voight, 1988), has emerged as a cornerstone of modern eruption forecasting efforts. This framework leverages accelerating precursory signals, such as seismicity or ground deformation, to estimate the timing of volcanic rupture, offering a systematic way to interpret transient unrest patterns.

Several mechanisms have been proposed to explain such precursory power law dynamics frequently observed in reawakening volcanoes. One mechanism is based on an analogy between catastrophic failures and critical points, highlighting the scaling symmetry of power laws, akin to critical phase transitions (Sornette, 2006). Another mechanism attributes the presence of power law acceleration to stress corrosion-induced subcritical crack growth and damage accumulation (Cornelius and Scott, 1993; Kilburn, 2003; Kilburn and Voight, 1998; Main, 2000). This failure forecast method, including its stochastic variant, has been applied to forecast the eruption timings of numerous volcanoes (Bevilacqua et al., 2022; Boué et al., 2016; Chastin and Main, 2003; Cornelius and Voight, 1994, 1995; Tárraga et al., 2008; Voight and Cornelius, 1991). However, despite its utility, the FFM approach has faced persistent limitations, as demonstrated by an extensive body of work (Bell et al., 2011a, 2011b, 2013). A critical limitation stems from the episodic rupture dynamics of reawakening volcanic systems (Acocella et al., 2023), which frequently exhibit non-linear, discontinuous inflation-deflation episodes prior to eruption. This behaviour deviates fundamentally from the gradual, monotonic acceleration assumed by classical power law models, complicating the detection of clear precursory trends.

The Log-Periodic Power Law Singularity (LPPLS) model extends the conventional power law framework by incorporating log-periodic components, enabling it to capture the intermittent oscillations of heterogeneous systems undergoing progressive failure toward global rupture (Anifrani et al., 1995; Sornette and Sammis, 1995). Our recent study has demonstrated that the LPPLS model outperforms conventional power law models, as evidenced by a comprehensive analysis of a global dataset covering landslides, rockbursts, glacier breakoffs, and volcanic eruptions (Lei and Sornette, 2024). By capitalising on the irregular and intermittent nature of geomaterial rupture, the LPPLS model exploits unsteady, non-monotonic signals—traditionally often regarded as noise—to refine its predictions, offering a promising framework for forecasting catastrophic events. Indeed, noise often reflects unexplored complexity, but with expanded analytical frameworks like LPPLS, it can reveal discernible patterns that were previously undetectable using incomplete models.

This Letter aims to present empirical evidence and theoretical insights into the statistical significance of log-periodicity in reawakening volcanoes, which has important implications for understanding and forecasting volcanic eruptions. More specifically, first, log-periodic patterns indicate the spontaneous hierarchical organisation of damage in heterogeneous crustal systems, offering valuable insights into the underlying mechanisms driving volcanic unrest; second, log-periodicity can improve the reliability of failure time prediction by constraining the model fit to the accelerating oscillatory pattern of volcanic activity.

The rest of the Letter is organised as follows. Section 2 introduces the LPPLS model along with the parametric calibration and nonparametric test methods. Section 3 presents the results of our analysis based on a comprehensive dataset of 34 historical eruption events. Section 4 discusses the significance of log-periodicity in volcanoes, interprets the possible underlying mechanisms, and explores the implications for eruption forecasting. Finally, section 5 concludes the paper.

## 2. Model description

The pre-eruptive unrest of a volcano is typically modelled by a nonlinear dynamic equation (Voight, 1988, 1989), expressing a power law relation between the first and second derivatives of an observable quantity Ω of the volcanic system (e.g. displacement, strain, tilt, energy release, earthquake count, and gas emission, etc.):

$$\ddot{\Omega} = \mu \dot{\Omega}^{\alpha}\text{, with } \alpha > 1, \tag{1}$$

where $\mu$ is a positive constant and $\alpha$ is an exponent characterising the nonlinearity of the accelerating dynamics. With $\alpha > 1$, positive feedbacks dominate, driving a super-exponential dynamic that culminates in a finite-time singularity (Johansen and Sornette, 2000; Main, 1999). This singular behaviour becomes evident if we integrate equation (1), yielding:

$$\dot{\Omega} = \zeta (t_c - t)^{-\beta}, \tag{2}$$

where $\zeta = (\beta/\mu)^{\beta}$, $\beta = 1/(\alpha-1)$, with $\beta > 0$ (for $\alpha > 1$) ensuring the emergence of a singularity as time $t$ approaches the critical time $t_c$. Further integrating equation (2) yields the power law time-to-failure model (Main, 1999; Voight, 1989, 1988):

$$\Omega(t) = A + B(t_c - t)^m\text{, with } m < 1, \tag{3}$$

where $m = 1-\beta = (\alpha-2)/(\alpha-1)$ is the singularity exponent, and $A$ and $B = -\zeta/m$ are constants. This power law relation exhibits continuous scale invariance, where scaling $t_c-t$ by an arbitrary factor $\lambda$ leads to a corresponding scaling of the observable (for $m \le 0$) or of the difference of the observable to its final value $A$ (for $0 < m < 1$) by the factor $\lambda^m$ which is independent of $t_c-t$.

We now extend the power law model by allowing the singularity exponent to take complex values, $m+i\omega$, which are generalised solutions of renormalisation group equations near critical points (Saleur et al., 1996) and naturally arise in dissipative systems (like volcanoes) with out-of-equilibrium dynamics and quenched disorder (Saleur and Sornette, 1996). The first-order Fourier expansion of the general solution of Ω leads to the LPPLS model (Anifrani et al., 1995; Lei and Sornette, 2024; Sornette and Sammis, 1995), given by:

$$\Omega(t) = A + \left\{B + C\cos\left[\omega \ln(t_c - t) - \phi\right]\right\}(t_c - t)^m\text{, with } m < 1. \tag{4}$$

where, compared to equation (3), three new parameters are involved, viz. a constant $C$, an angular log frequency $\omega$, and a phase shift $\phi$. This introduces a log-periodic correction with a relative amplitude of $C/B$ (typically on the order of 10%) superimposed on the overal power law trend with a pre-factor $B$. In this case, continuous scale invariance is partially broken into discrete scale invariance (Sornette, 1998), where the observable remains invariant under scaling of $t_c–t$ by integer powers of a fundamental scaling factor $\lambda > 1$ that is related to $\omega$ through $\omega = 2\pi/\ln\lambda$. The local maxima of the log-periodic term in the LPPLS formula thus form a geometric time series $\{t_1, t_2, \ldots, t_k, \ldots\}$, satisfying $t_c–t_k = \lambda^{-k}\exp(\phi/\omega)$, where $k$ is an integer, such that the argument of the cosine function in equation (4) is an integer multiple of $2\pi$. By inherently encoding this discrete hierarchy of time scales, the LPPLS model can capture the intermittent dynamics of rupture phenomena, where burst frequency increases geometrically towards $t_c$. This pattern reflects the localised and threshold nature of the mechanics of rupture in heterogeneous materials (Johansen and Sornette, 2000; Sornette, 2002). Note that only the first-order correction is included in equation (4), while higher-order terms with smaller amplitudes also exist but are generally less significant (Zhou and Sornette, 2002a).

We employ a stable and robust parametric calibration scheme to calibrate the LPPLS model against time series data, with the procedures briefly summarised as follows (see Supplementary Text S1 for more detailed descriptions). First, we use the Lagrange regularization method (Demos and Sornette, 2019) to identify the onset time $t_0$ of the volcanic unrest, whose endpoint is set at the last available data point before the eruption. Then, the optimal set of LPPLS model parameters is estimated using the ordinary least squares method, through a minimisation of the sum of the squares of the differences between the model prediction and the observational data (Filimonov and Sornette, 2013).

Additionally, we conduct a nonparametric test to qualify the presence of log-periodic oscillations in the time series data. We apply the following transformation (Johansen and Sornette, 2001) to remove the background power law trend:

$$\varepsilon(t) = \frac{\Omega(t) - A - B(t_c - t)^m}{C(t_c - t)^m}, \tag{5}$$

where the normalised residual $\varepsilon$ as a function of the log normalised time $\tau = (t_c–t)/(t_c–t_0)$ should obey a pure cosine function, i.e. $\cos(\omega\ln\tau–\phi)$, if equation (4) gives a perfect description of the data. In most cases, the $\varepsilon$ signals would be unevenly spaced on the logarithmic time $\ln\tau$, as occurs for instance when they are evenly sampled on the linear time scale $\tau$, rendering the standard fast Fourier transform unsuitable for our analysis. To address this issue, we use the Lomb spectral method (Lomb, 1976), which performs least-squares fitting of sinusoids to local data points, with no requirement of equidistant sampling (see Supplementary Text S2), making it ideal for our problem of detecting periodicity on the logarithmic time to a singularity.

## 3. Results

The first case to study is Axial Seamount, which is an active volcano with a summit caldera at approximately 1.5 km depth and a base at around 2.4 km, located in the Pacific Ocean and about 500 km off the coast of Oregon, USA. This basaltic shield volcano, fed by magma from the Cobb hotspot on the Juan de Fuca spreading ridge, has experienced three eruptive episodes over the past 30 years, specifically in 1998, 2011, and 2015 (Chadwick et al., 2022). A large shallow magma reservoir may lie approximately 1.5 to 2.5 km beneath the caldera, accompanied by a series of deeper, stacked sills at depths ranging from 2.5 to 4.5 km (Carbotte et al., 2020). Since 1998, this submarine volcano has been closely monitored by various seafloor instruments and since 2014 by a cabled observatory, where the seafloor vertical displacement is measured by bottom and mobile pressure recorders at a resolution of ~1 cm (Kelley et al., 2014). We focus on the eruption event that occurred in April 2011, during which a total volume of about 99 million $m^3$ lava was erupted into the sea (Chadwick et al., 2012). This volcano displayed a series of short-term inflation-deflation cycles over about 8 years prior to the 2011 eruption, resulting in an accelerating and oscillating evolutionary pattern of seafloor deformation, which is effectively captured by the LPPLS model (Fig. 1a). Log-periodic oscillations are also evident in the $\epsilon$-ln$\tau$ plot (inset of Fig. 1a), despite considerable fluctuations around the overall sinusoidal trend, which may arise from the underlying complex dynamics and/or the presence of various disturbances. From the Lomb periodogram (Fig. 1b), we can identify a dominant peak at the log frequency $f \approx 1.52$ (with the angular log frequency $\omega \approx 9.55$ and the scaling ratio $\lambda \approx 1.93$). We have also performed analysis of the seafloor deformation data for the other two eruption events at this volcano. Despite the limited observational data for the 1998 eruption and the short acceleration period preceding the 2015 eruption, the results yield a similar value of dominant frequency at $f \approx 1.2$ with $\omega \approx 7.7$ and $\lambda \approx 2.2$ (see Fig. S1 in Supplementary Materials), which is compatible with that for the 2011 eruption.

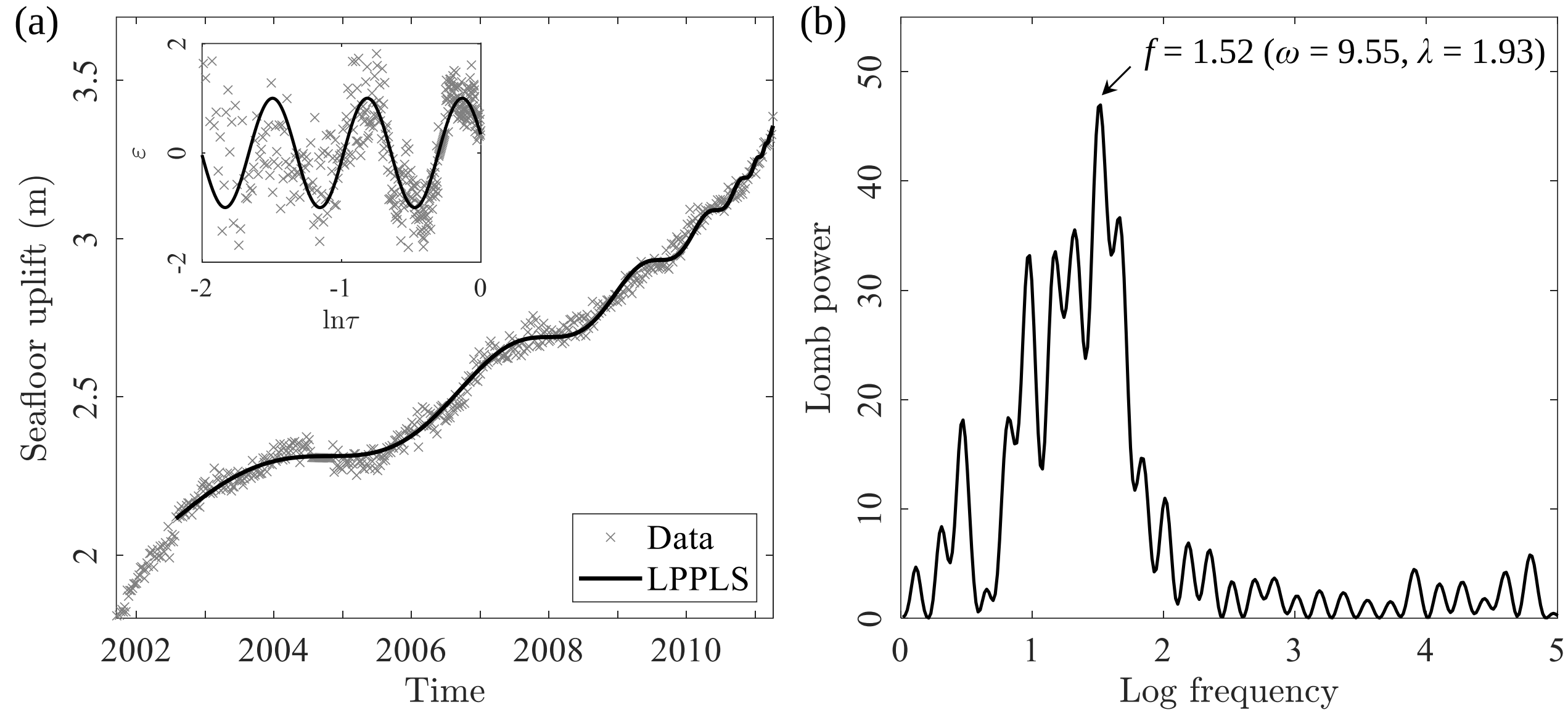

**Fig. 1.** (a) Monitoring data showing the temporal evolution of seafloor uplift (aggregated on a weekly basis) at the Axial Seamount volcano in the Pacific Ocean, prior to an eruption in April 2011. The raw data originally recorded by a single-station bottom pressure recorder at very high frequencies (i.e. every 15 seconds) are aggregated on a much lower frequency (i.e. weekly) to expedite the model calibration and facilitate the visualisation. Inset displays the variation of normalised residual $\varepsilon$ as a function of log normalised time $\tau = (t_c - t)/(t_c - t_0)$. (b) The Lomb periodogram analysis for detecting log-periodic oscillatory components in the data.

The second example is Sierra Negra, which is a shield volcano located at the southeastern end of Isabela Island, Ecuador. The large summit caldera of this volcano, with dimensions of 7.5 km by 9.5 km and a depth of around 100 m, may be underlain by a flat-topped, sill-like magma reservoir at approximately 2 km depth beneath it (Yun et al., 2006). This active basaltic volcano has erupted three times over the last 30 years: in 1979, 2005, and 2018. Since 2002, a local network of continuous Global Positioning System stations has been in place, with the pre-eruptive and co-eruptive ground deformation captured for the 2005 and 2018 events (Bell et al., 2021; Chadwick et al., 2006), during which approximately 150 million $m^3$ and 140 million $m^3$ of lava were erupted, respectively (Geist et al., 2008; Vasconez et al., 2018). The data revealed that this volcano experienced significant surface uplift during the pre-eruptive inflation phase prior to both 2005 and 2018 eruptions (Bell et al., 2021; Chadwick et al., 2006). The existence of log-periodicity in Sierra Negra is demonstrated by the excellent fit of the LPPLS model to the time series of surface uplift and the sinusoidal-like signals in the $\epsilon$-ln$\tau$ plot (Fig. 2a and c). The Lomb periodograms reveal a dominant peak at $f = 1.51$ and 1.55 for the 2005 and 2018 eruptions, respectively, with $\omega$ = 9.51 and 9.74, and $\lambda$ = 1.94 and 1.91 correspondingly (Fig. 2b and d).

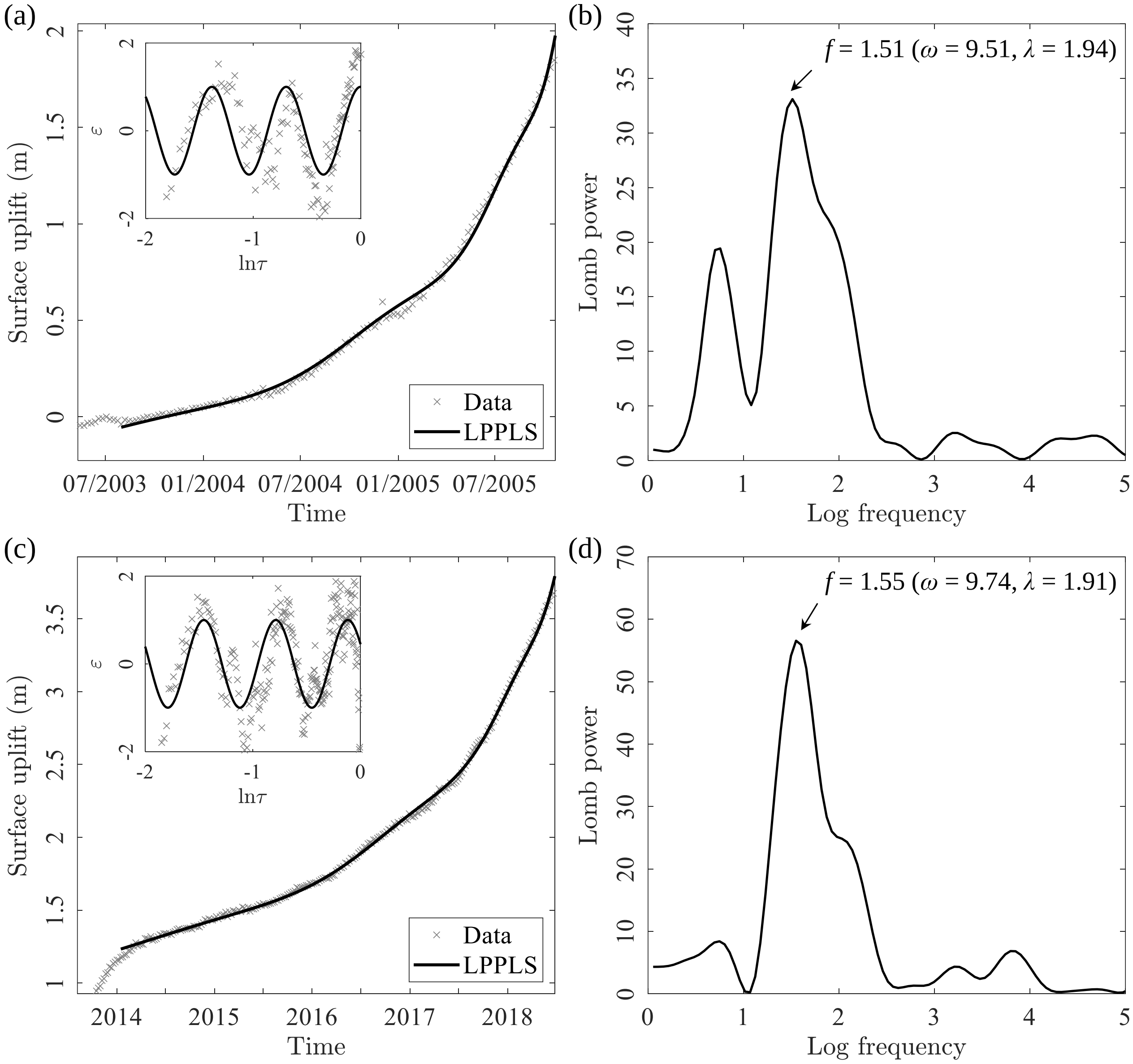

**Fig. 2.** Temporal evolution of surface uplift recorded by the continuous Global Positioning System stations GV02 and GV04 at the Sierra Negra volcano in Ecuador, prior to the (a) 2005 and (b) 2018 eruptions. The raw data originally recorded daily are aggregated on a weekly basis to expedite the model calibration and facilitate the visualisation. Insets in the left panel display the variation of normalised residual $\varepsilon$ as a function of log normalised time $\tau = (t_c - t)/(t_c - t_0)$. The Lomb periodogram analysis for detecting log-periodic oscillatory components in the pre-eruptive deformational pattern of the (b) 2005 and (d) 2018 events.

The third study case is the Mount St. Helens, which is an active stratovolcano located in the Pacific Northwest region of the United States. Between 1980 and 1986, the volcano underwent a series of explosive eruptions and dome-building events, resulting in the formation of a new lava dome within the crater. Its activity has been continuously monitored using various instruments (Chadwick et al., 1983; Dzurisin et al., 1983; Malone et al., 1983), capturing significant deformational and seismic precursors associated with pre-eruptive dome inflation driven by magma intrusion. We focus on an eruption event that occurred in March 1982, which was preceded by distinct accelerating and oscillatory patterns in earthquake and tiltmeter measurements (Fig. 3a, c, and e). The existence of log-periodicity in the earthquake count data is demonstrated by the LPPLS fit and the $\epsilon$-ln$\tau$ pattern (Fig. 3a), with the Lomb periodogram yielding $f = 0.77$, $\omega = 4.85$, and $\lambda = 3.65$ (Fig. 3b). The LPPLS fit and $\epsilon$-ln$\tau$ plot for the tilt data also confirms the presence of log-periodicity (Fig. 3c and e), and the Lomb spectral analysis indicates $f = 0.99$, $\omega = 6.23$, and $\lambda = 2.74$ for the tilt along the radial direction (Fig. 3d) and $f = 1.16$, $\omega = 7.26$, and $\lambda = 2.37$ for the tilt along the radial direction (Fig. 3f). Notably, the $\lambda$ values derived from tiltmeter measurements are significantly higher than that obtained from earthquake counts.

The next case study focuses on Merapi volcano in Indonesia, one of the most active and hazardous volcanoes globally. It has erupted regularly, about every 4 to 6 years, typically with low explosivity and involving the formation of a lava dome. A major eruption took place in June 2006, releasing approximately 5.3 million $m^3$ of magma, just before the Mw 6.4 Yogyakarta earthquake (Ratdomopurbo et al., 2013). The existence of log-periodicity is indicated by the LPPLS fit to the data and the cyclical pattern in the $\epsilon$-ln$\tau$ plot (Fig. 4a). On the Lomb periodogram, a dominant peak is found at $f = 0.78$ (with $\omega = 4.90$ and $\lambda = 3.61$), while a harmonic at 1.61 is also noticeable, corresponding approximately to the second harmonic $2f$ (Fig. 4b). In October 2010, Merapi exploded again but in a much more violent manner, marking its largest eruption over a century (Surono et al., 2012). This event was preceded by a clear accelerating and oscillating pattern in the earthquake count data, which is well captured by the LPPLS model (Fig. 4c). The $\epsilon$-ln$\tau$ plot also confirms the presence of log-periodicity (inset of Fig. 4c), and the Lomb spectral analysis indicates $f = 1.42$, $\omega = 8.91$, and $\lambda = 2.02$ (Fig. 4d). Notably, the Lomb periodogram highlights a series of harmonics occurring at integer multiples of this fundamental frequency $f$—another signature of log-periodicity.

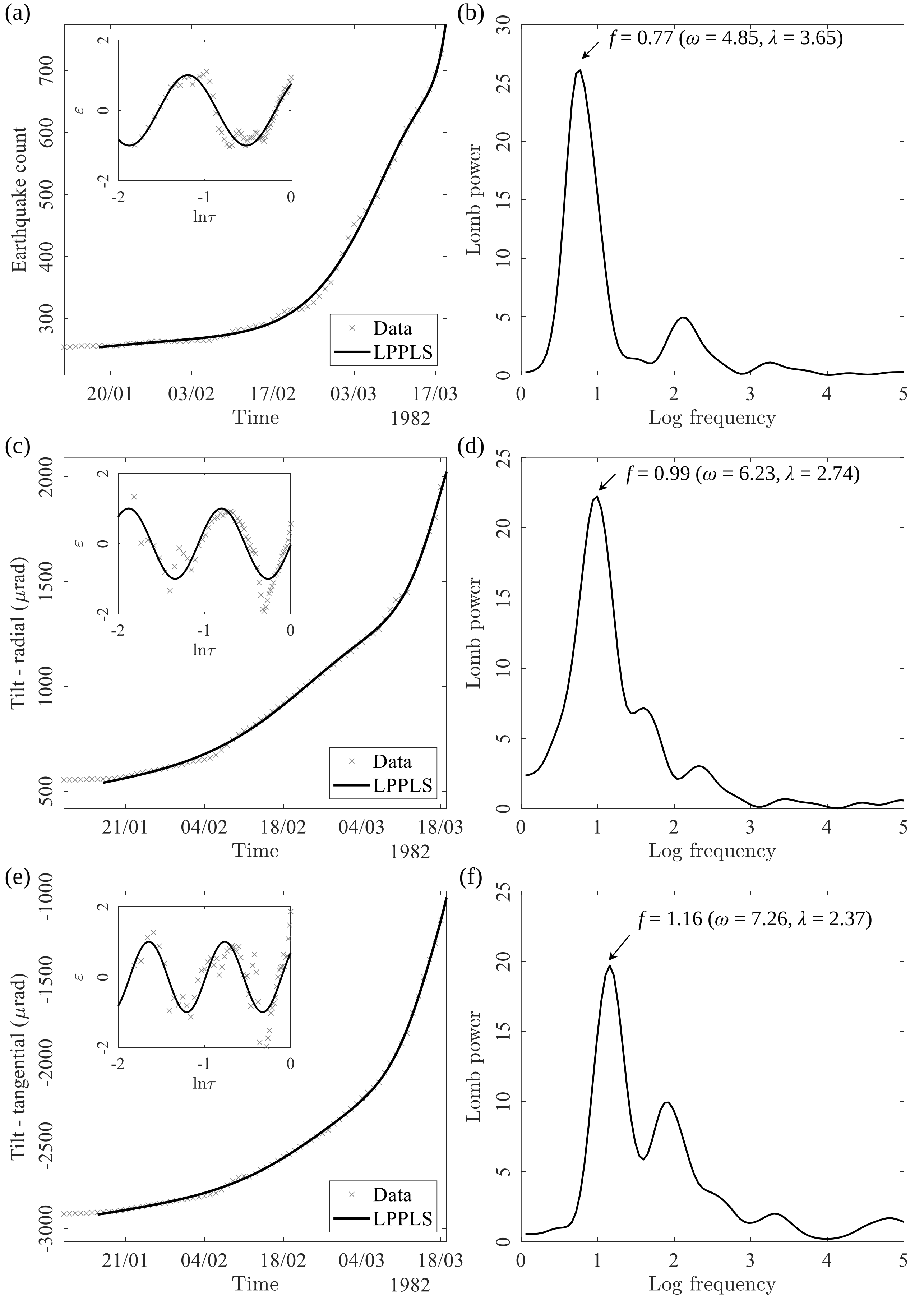

**Fig. 3.** Time series of earthquake count and tiltmeter measurements at the Mount St. Helens volcano

in the United States, prior to an eruption in March 1982. Insets in the left panel display the variation of normalised residual $\varepsilon$ as a function of log normalised time $\tau = (t_c - t)/(t_c - t_0)$. The Lomb periodogram analysis for detecting log-periodic oscillatory components in the (b) seismic data as well as (d) radial and (f) tangential tiltmeter measurements.

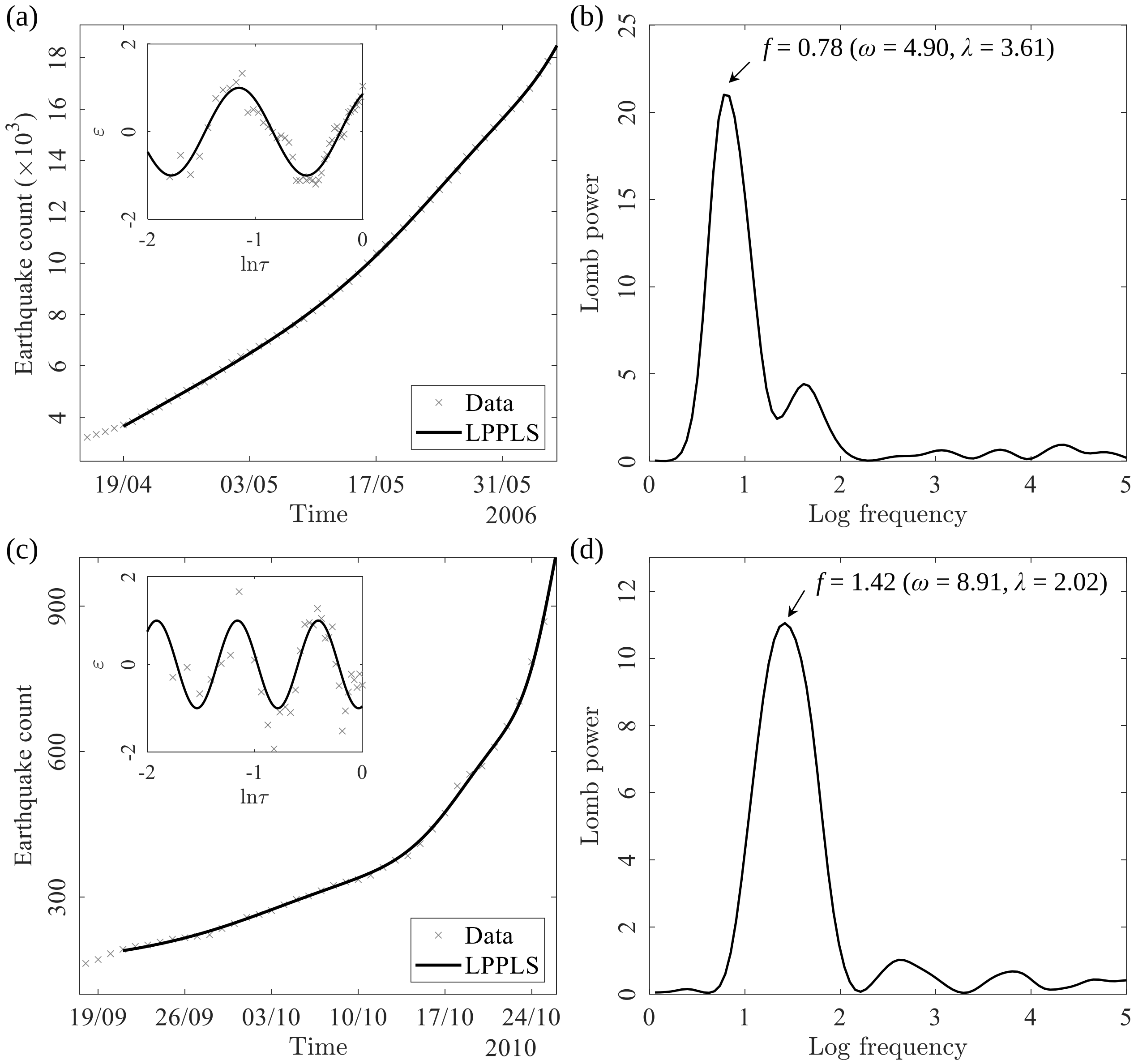

**Fig. 4.** Time series of earthquake count at the Merapi volcano in Indonesia, prior to the (a) 2006 and (c) 2010 eruptions. Insets in the left panel display the variation of normalised residual $\varepsilon$ as a function of log normalised time $\tau = (t_c - t)/(t_c - t_0)$. The Lomb periodogram analysis for detecting log-periodic oscillatory components in the data for the (b) 2006 and (d) 2010 eruptions.

We further show our analysis on two other representative volcanoes, i.e. Mount Pinatubo in Philippines and Mount Etna in Italy. In June 1991, Pinatubo experienced a massive cataclysmic explosion and discharged about 5 billion $m^3$ pyroclastic material (Kress, 1997), making it the second-largest terrestrial eruption in the 20th century. A clear log-periodic oscillatory pattern decorating an overall power law acceleration can be seen in the time series of earthquake count which is well captured by the LPPLS model (Fig. 5a). The significance of log-periodic oscillations is indicated by the sinusoidal-like signals in the $\epsilon$-ln$\tau$ plot (inset of Fig. 5a) and the emergence of a major Lomb peak (Fig. 5b), pointing to $f$ = 1.06, $\omega$ = 6.64, and $\lambda$ = 2.58. Regarding the Etna volcano that erupted in September 1989 (Rymer et al., 1993), the existence of log-periodicity is also demonstrated by the excellent fit of the LPPLS model to the time series data of earthquake count and the visible cyclical pattern in the $\epsilon$-ln$\tau$ plot (Fig. 6a). On the Lomb periodogram (Fig. 6b), a major peak is observed at $f$ = 1.87 corresponding to $\omega$ = 11.77 and $\lambda$ = 1.71, with a secondary peak also identified at the log frequency of 3.65 indicating the second harmonic.

In total, we have performed the LPPLS calibration and Lomb periodogram analysis on a large dataset including 34 volcanic eruptions (Supplementary Table 1 and Fig. S2-S4). This dataset includes different types of volcanoes (stratovolcano, shield volcano, and complex volcano) monitored by different instruments (e.g. electronic tiltmeter, seismic/geochemical stations, bottom/mobile pressure recorders, and Global Positioning System) (Supplementary Table 1). In Fig. 7, we show the histograms of selected key parameters derived from the Lomb spectral analysis of this large dataset (see Supplementary Table S2 for the parameter values of each eruption event). The angular log frequency $\omega$ spans from 3.5 to 14, with a notable concentration between 4 and 10 (Fig. 7a). Correspondingly, the scaling ratio $\lambda$ mostly falls in the range of 1.5 to 6, exhibiting a concentration around 2 (Fig. 7b). The frequency distributions of $\omega$ and $\lambda$ obtained using the Lomb method generally align with those derived from the LPPLS calibration (Supplementary Fig. S5). The frequency and cumulative distributions of maximum Lomb peak heights $P_{\max}$ of these eruptions indicates that 94% exceed 5, 79% are beyond 10, and 59% reach over 20 (Fig. 7c), highlighting the significance of log-periodicity (Zhou and Sornette, 2002b). The first-to-second peak ratio $\eta$, which quantifies the relative strength of the two highest peaks in each Lomb periodogram, reveals that 62% of these events are beyond 2, 47% surpass 3, and 32% exceed 4 (Fig. 7d). This reinforces the evidence for log-periodicity, as it highlights the presence of both a fundamental log frequency and its first harmonic.

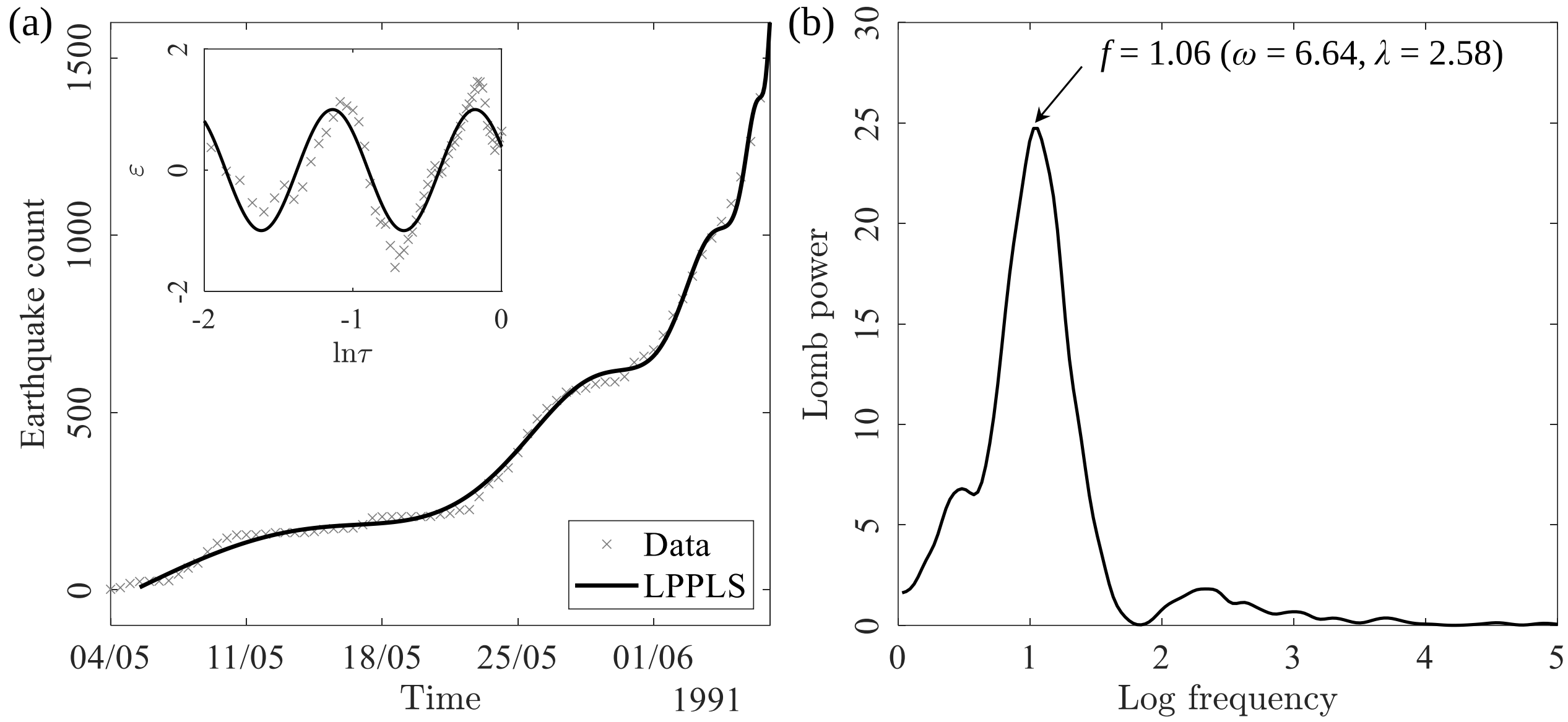

**Fig. 5.** (a) Time series of earthquake count at the Pinatubo volcano in Philippines, prior to the 1991 eruption. Inset displays the variation of normalised residual $\varepsilon$ as a function of log normalised time $\tau = (t_c - t)/(t_c - t_0)$. (b) The Lomb periodogram analysis for detecting log-periodic oscillatory components in the data.

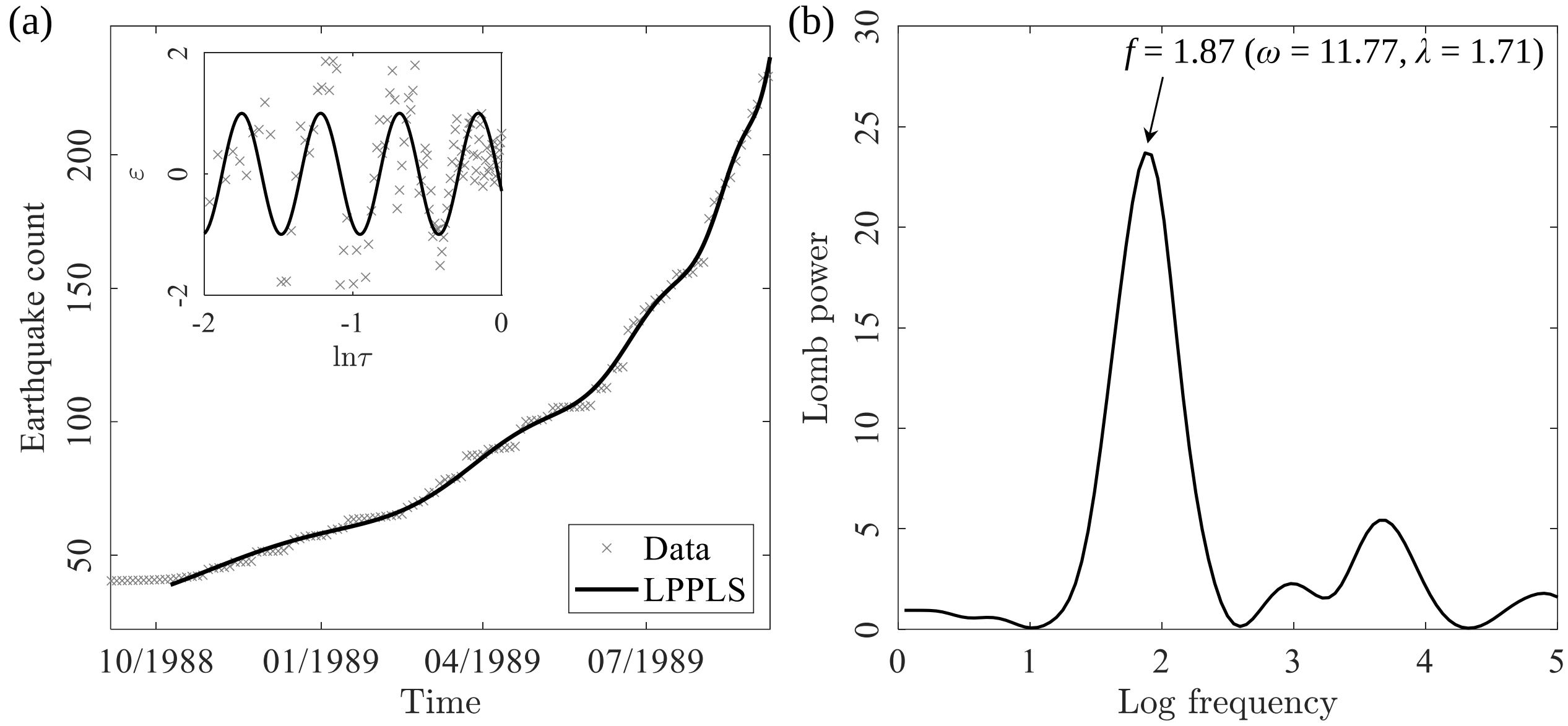

**Fig. 6.** (a) Time series of earthquake count at the Etna volcano in Italy, prior to the 1989 eruption. Inset displays the variation of normalised residual $\varepsilon$ as a function of log normalised time $\tau = (t_c - t)/(t_c - t_0)$. (b) The Lomb periodogram analysis for detecting log-periodic oscillatory components in the data.

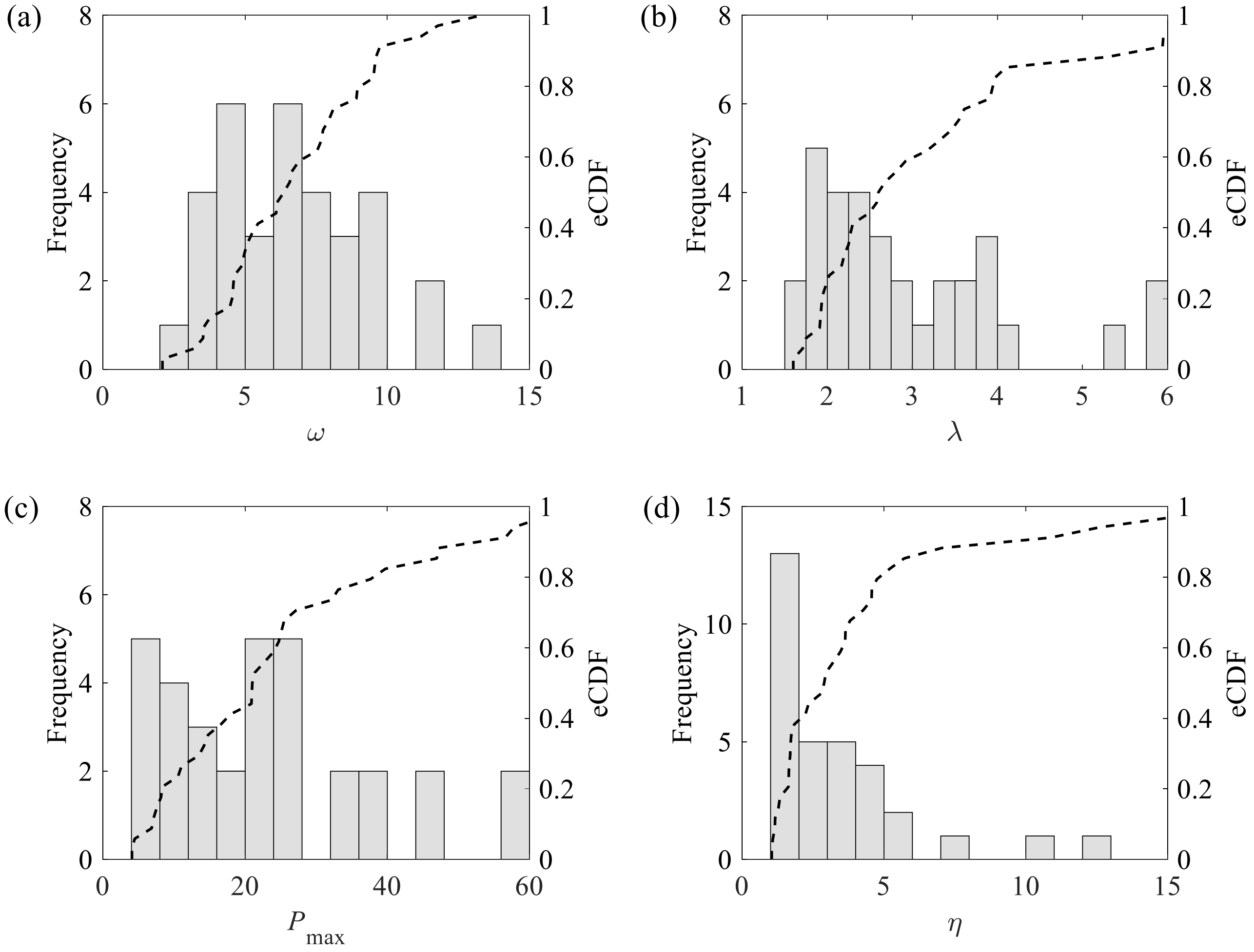

**Fig. 7.** Histograms together with the empirical cumulative distribution function (eCDF) of the parameter values derived from the Lomb periodogram analysis of 34 volcanic eruptions, including (a) the log-periodic frequency $\omega$, (b) the scaling ratio $\lambda$, (c) the maximum Lomb peak height $P_{max}$, and (d) the first-to-second peak ratio $\eta$.

## 4. Discussion

Our findings suggest that log-periodicity is a common precursor to volcanic eruption, demonstrated by the strong agreement between the LPPLS model and monitoring data from a diverse range of volcanoes (Figs. 1-6 and Supplementary Figs. 1-4). It is important to note that the possibility of overfitting by the LPPLS model has been ruled out in our previous study based on Akaike and Bayesian information criteria (Lei and Sornette, 2024). Our LPPLS parametric calibration reveals that the singularity exponent $m$, which characterises the nonlinearity of power law acceleration dynamics, predominantly ranges from -1 to 1 with a concentration between 0.5 and 1. Correspondingly, the exponent $\alpha = 1+1/(1-m)$ is mostly larger than 1.5, with a concentration in the range above 3. This deviates from the commonly assumed $\alpha$ = 2.0 in the inverse rate method (Cornelius and Voight, 1995, 1994; Voight, 1989, 1988; Voight and Cornelius, 1991), suggesting that special caution is needed when adopting this assumption in practice. The transition from the power law framework of the FFM model to the LPPLS model yields a substantial improvement in fitting volcanic unrest time series, as evidenced by a marked reduction in residual variance (Lei and Sornette, 2024). Crucially, the statistical robustness of log-periodic components is underscored by oscillatory patterns in FFM residuals when plotted on a logarithmic time axis nearing the critical time. These coherent and stable oscillations—incompatible with stochastic noise or power law scaling alone—provide diagnostic validation of the LPPLS model's ability to capture the intermittent bursty behaviors in the accelerated dynamics in all the eruption events studied (see insets of Figs. 1-6 and Supplementary Figs. 1-4). Some discrepancies in a few cases may arise from either the scarcity of available data or the effect of higher-order harmonics that are not included in this version of the LPPLS model, which includes only the first-order log-periodic correction to the pure power law. Higher-order corrections could be incorporated (Gluzman and Sornette, 2002; Johansen and Sornette, 2001), which however would make the calibration process less parsimonuous. The relative amplitude of the log-periodic components typically ranges from 0.01 to 0.2, with a concentration around 0.1, consistent with the hypothesis that it is on the order of $10^{-1}$ for systems approaching failure (Sornette, 1998). Our nonparametric tests offer additional support for the presence of log-periodicity in volcanic activity. In particular, the high Lomb peaks across different eruption events, with most surpassing 10 and over half exceeding 20 (Fig. 7c), reveal the statistical significance of log-periodic components (Zhou and Sornette, 2002b). If the data were associated with Gaussian noise, we could analytically calculate the false-alarm probability, which quantifies the chance that random noise is mistakenly identified as a genuine log-periodic signal (Supplementary Text S2). Among the 34 events studied, only 3 show a false-alarm probability exceeding 0.05 (Supplementary Table S2). Although the noise in the real data is likely non-Gaussian, the high first-to-second Lomb peak ratios, with most exceeding 2 and half surpassing 3, still provide compelling evidence of log-periodicity (Zhou and Sornette, 2002b). This is further supported by the high signal-to-noise ratios in most cases, with 82% above 1, 56% exceeding 1.5, and 38% surpassing 2 (Supplementary Table S2), a metric that is applicable for different types of noise (Zhou and Sornette, 2002a, 2002b). It is worth mentioning that log-periodicity may also arise due to artifacts (Huang et al., 2000), but for most cases that we analysed, the most probable artificial log frequency is very different from the detected log frequency in the Lomb peridogram (Supplementary Table S2). In some cases, they are close but with high signal-to-noise ratios that could still exclude the possible noise origin of the detected log-periodicity (Supplementary Table S2).

The log-periodic patterns provide important clues about the possible underlying mechanisms behind the observed intermittent rupture behaviour of volcanoes. Log-periodicity indicates the presence of discrete scale invariance (Saleur et al., 1996; Sornette, 1998), linked to complex critical exponents that are commonly observed in nonunitary (dissipative) systems with quenched disorder exhibiting out-of-equilibrium dynamics (Saleur and Sornette, 1996). One possible mechanism for discrete scale invariance is a cascade of ultraviolet Mullins-Sekerka instabilities that occurs in the growth processes of diffusion-limited aggregation and crack array propagation, with a preferred scaling ratio close to 2 (Huang et al., 1997; Johansen and Sornette, 1998; Sornette, 1998; Sornette et al., 1996). This may explain the frequent occurrence of $\lambda$ around 2 (Fig. 7b). In the context of volcanoes, diffusion-limited aggregation can be envisioned as the flow of magma through porous media between the magma chamber and the surrounding crust. This mechanism may explain the observed $\lambda \approx 2$ for the Axial Seamount volcano (see Fig. 1b), aligning with previously proposed conceptual models for the short-term inflation-deflation cycles at this volcano (Chadwick et al., 2022). Similarly, log-periodicity could spontaneously emerge in crack propagation, where larger cracks experience less screening and grow more quickly, while smaller cracks are inhibited due to stress shadowing and crack interactions (Huang et al., 1997). This theory, formulated for a system of parallel cracks, may account for the observed $\lambda \approx 2$ for the Sierra Negra volcano, where dyke propagation and fissure opening have been observed in the northern sector of the caldera (Rezaeifar et al., 2024) and interpreted to relate to the seismic tremors recorded (Li et al., 2022). Another possible mechanism involves the interplay between stress drop during seismic phases and stress corrosion during inter-seismic phases, with a mean-field prediction of $\lambda = 3.6$ (Lee and Sornette, 2000). This theory, which may explain some of the volcanoes having $\lambda$ values around 3.5-4.0 (Fig. 7b), is consistent with the suspected involvement of pre-existing fault reactivation and subcritical crack growth in triggering volcanic eruptions (Gudmundsson, 2016; Heap et al., 2011; Kilburn, 2003; Kilburn and Voight, 1998).

Additionally, log-periodicity could emerge from the interplay of inertia, damage, and healing (Ide and Sornette, 2002; Sornette and Ide, 2003), which may explain the large variability in $\lambda$ values observed across different volcanoes, reflecting the differences in their healing properties. In the context of volcanoes, the presence of inertia, damage, and healing is supported by the commonly observed volcano-tectonic earthquakes, dyke intrusion, and magma cooling/crystallisation (Acocella et al., 2023; Caricchi et al., 2021).

Thus, discrete scale invariance can emerge spontaneously through mechanisms such as (i) cascades of instabilities, (ii) the interplay between stress drop during seismic phases and stress corrosion during inter-seismic phases, or (iii) the interplay of inertia, damage, and healing. In addition, discrete scale invariance derives naturally in systems where the structure itself exhibits hierarchical organisation, such as fracture networks governed by discrete preferred scaling ratios in the uppermost crust, which imprint log-periodic signatures (Sornette, 1998). These various mechanisms are likely to coexist and interact in actual volcanoes, with the dominant mechanism changing across different sites or over time at the same site. For example, the Merapi volcano exhibited $\lambda \approx 3.6$ during the pre-eruptive inflation before the 2006 eruption (Fig. 4b), indicating a possible dominance of stress drop along geological structures and stress corrosion damage in crustal rocks, while $\lambda \approx 2$ during the 2010 unrest (Fig. 4d) may suggest a dominant role of dyke intrusion and/or magma diffusion. Interestingly, for the 1982 eruption at Mount St. Helens, the $\lambda$ value derived from earthquake counts is around 3.6 while that from tiltmeter measurements is about 2.5, which may be attributed to the fact that they reflect distinct processes (e.g. faulting versus dyking, respectively) operating within the same system.

In addition to shedding light on the mechanisms driving volcanic unrest, log-periodic signals could be useful for forecasting impending volcanic eruptions in practice. More specifically, by "locking" into the oscillatory temporal evolution pattern of reawakening volcanoes, the LPPLS model can transform the information of intermittency from traditionally perceived noise into valuable constraints to improve the prediction of the critical time of failure (Lei and Sornette, 2024; Sornette, 2002). Its potential for prospective forecasting of volcanic eruptions will be explored in our future work. Moreover, log-periodic signatures could serve as indicators for early warning, a concept proven effective for forecasting financial crises (Demirer et al., 2019) and to be explored for volcanoes in our future research. However, the Lomb nonparametric analysis alone may be less robust for prediction than the LPPLS parametric fit and finds its use as a diagnostic tool (Zhou and Sornette, 2002a). Finally, it is important to note that the presence of log-periodic signatures in a volcano only indicates that the system is at or close to the critical point and does not guarantee the definite occurrence of an eruption. In other words, log-periodicity alone is insufficient to distinguish between volcanic crises that will escalate into an eruption and those that will not. Many cases of significant volcanic unrest have not resulted in eruptions (Phillipson et al., 2013), while there are also situations where eruptions occurred without clear precursory accelerating signals (Smittarello et al., 2022). Therefore, log-periodicity should be combined with other geophysical and geochemical indicators to improve the reliability of eruption forecasts.

## 5. Conclusions

To conclude, through comprehensive parametric and nonparametric tests on a large dataset of 34 volcanic eruptions, we provided strong empirical evidence, supported by solid theoretical arguments, demonstrating the statistical significance of log-periodic oscillations in reawakening volcanoes. Log-periodicity seems to be a ubiquitous feature decorating the power law finite-time singularities during pre-eruptive volcanic unrest. It reveals discrete scale invariance in volcanic activity that is inherent to the intermittent dynamics of damage and rupture processes in heterogeneous systems. Drawing from the log-periodic patterns derived from empirical data, we inferred that log-periodicity in volcanoes may arise from various mechanisms, such as diffusion-dominated magma flow, magma-driven dyke intrusion, interaction between stress drop and stress corrosion, and/or interplay of inertia, damage, and healing, within magmatic and crustal systems. Our findings have important implications for volcano forecasting, since log-periodicity can help turn volcanic intermittency from traditionally perceived noise into valuable information to constrain predictions.

**CRediT authorship contribution statement**

QL: Conceptualisation; Methodology; Software; Validation; Formal analysis; Investigation; Resources; Writing - Original Draft; Visualization; Project administration.

DS: Conceptualisation; Methodology; Software; Resources; Writing - Review & Editing; Funding acquisition.

**Declaration of competing interest**

The authors declare that they have no known competing financial interests or personal relationships that could have appeared to influence the work reported in this paper.

**Acknowledgement**

Q.L. acknowledges the National Academic Infrastructure for Supercomputing in Sweden (NAISS), partially funded by the Swedish Research Council through grant agreement no. 2022-06725, for awarding this project access to the LUMI supercomputer, owned by the EuroHPC Joint Undertaking and hosted by CSC (Finland) and the LUMI consortium. D.S. acknowledges partial support from the National Natural Science Foundation of China (Grant No. U2039202, T2350710802), from the Shenzhen Science and Technology Innovation Commission (Grant No. GJHZ20210705141805017) and the Center for Computational Science and Engineering at the Southern University of Science and Technology.

**Data availability**

No new data were produced in this work. The data underlying our study are either from prior published studies or datasets, with sources provided in the Supporting Information.

# Supplementary Materials for
# Log-Periodic Precursors to Volcanic Eruptions: Evidence from 34 Events

Qinghua Lei[1,*], Didier Sornette[2]

[1]*Department of Earth Sciences, Uppsala University, Uppsala, Sweden*

[2]*Institute of Risk Analysis, Prediction and Management, Academy for Advanced Interdisciplinary Studies, Southern University of Science and Technology, Shenzhen, China*

**Contents of this file**

**Text S1. Calibration of the LPPLS model**

We implement a robust calibration scheme (Demos and Sornette, 2019; Filimonov and Sornette, 2013) to calibrate the LPPLS model to time series data. Consider a time series $\mathbf{\Omega} = \{\Omega_1, \Omega_2, ..., \Omega_N\}$ measured at time stamps $\mathbf{t} = \{t_1, t_2, ..., t_N\} \in [t_0, t_\mathrm{f}]$, where $N$ is the total number of time stamps, and $t_0$ and $t_\mathrm{f}$ mark the start and final points of the time window for LPPLS fitting.

The original LPPLS formula is written as:

$$\Omega(t) = A + \left\{B + C\cos\left[\omega \ln(t_\mathrm{c} - t) - \phi\right]\right\}(t_\mathrm{c} - t)^m, \tag{S1}$$

with the parameter set $\mathbf{\theta} = \{A, B, C, t_\mathrm{c}, m, \omega, \phi\}$ having 7 parameters, among which the first 3 are linear and the last 4 are nonlinear. Introducing $C_1 = C\cos\phi$ and $C_2 = C\sin\phi$, equation (S1) is rewritten as:

$$\Omega(t) = A + B(t_\mathrm{c} - t)^m + C_1(t_\mathrm{c} - t)^m \cos\left[\omega \ln(t_\mathrm{c} - t)\right] + C_2(t_\mathrm{c} - t)^m \sin\left[\omega \ln(t_\mathrm{c} - t)\right], \tag{S2}$$

with the new parameter set $\mathbf{\theta} = \{A, B, C_1, C_2, t_\mathrm{c}, m, \omega\}$ having 7 parameters as well, among which however the first 4 are linear and the last 3 are nonlinear. Using the ordinary least squares method, the cost function is defined as:

* Corresponding author: qinghua.lei@geo.uu.se

$$F(\boldsymbol{\theta}_{\mathrm{LPPLS}};\boldsymbol{\Omega},\mathbf{t})=\sum_{i=1}^{N}\left\{\Omega_i-A-B(t_{\mathrm{c}}-t_i)^m-C_1(t_{\mathrm{c}}-t_i)^m\cos\left[\omega\ln(t_{\mathrm{c}}-t_i)\right]-C_2(t_{\mathrm{c}}-t_i)^m\sin\left[\omega\ln(t_{\mathrm{c}}-t_i)\right]\right\}^2, \quad \text{(S3)}$$

which will be minimised to estimate model parameters:

$$\hat{\boldsymbol{\theta}}=\arg\min_{\boldsymbol{\theta}}F(\boldsymbol{\theta};\boldsymbol{\Omega},\mathbf{t}). \quad \text{(S4)}$$

We enslave the 4 linear parameters $\{A,B,C_1,C_2\}$ to the 3 nonlinear ones $\{t_{\mathrm{c}},m,\omega\}$, so that the minimisation problem is reduced to:

$$\{\hat{t}_{\mathrm{c}},\hat{m},\hat{\omega}\}=\arg\min_{t_{\mathrm{c}},m,\omega}F_1(t_{\mathrm{c}},m,\omega), \quad \text{(S5)}$$

with

$$F_1(t_{\mathrm{c}},m,\omega)=\min_{A,B,C_1,C_2}F(A,B,C_1,C_2,t_{\mathrm{c}},m,\omega)=F(t_{\mathrm{c}},m,\omega,\hat{A},\hat{B},\hat{C}_1,\hat{C}_2). \quad \text{(S6)}$$

The 4 linear parameters $\{A,B,C_1,C_2\}$ can be estimated by solving the optimisation problem for fixed values of the 3 nonlinear parameters $\{t_{\mathrm{c}},m,\omega\}$:

$$\{\hat{A},\hat{B},\hat{C}_1,\hat{C}_2\}=\arg\min_{A,B,C_1,C_2}F(A,B,C_1,C_2,t_{\mathrm{c}},m,\omega), \quad \text{(S7)}$$

whose solution can be analytically derived from the system of linear equations below:

$$\begin{bmatrix} N & \sum f_i & \sum g_i & \sum h_i \\ \sum f_i & \sum f_i^2 & \sum f_i g_i & \sum f_i h_i \\ \sum g_i & \sum f_i g_i & \sum g_i^2 & \sum g_i h_i \\ \sum h_i & \sum f_i h_i & \sum g_i h_i & \sum h_i^2 \end{bmatrix}\begin{bmatrix}\hat{A}\\ \hat{B}\\ \hat{C}_1\\ \hat{C}_2\end{bmatrix}=\begin{bmatrix}\sum\Omega_i\\ \sum\Omega_i f_i\\ \sum\Omega_i g_i\\ \sum\Omega_i h_i\end{bmatrix}, \quad \text{(S8)}$$

where $f_i=(t_{\mathrm{c}}-t_i)^m$, $g_i=(t_{\mathrm{c}}-t_i)^m\cos\left[\omega\ln(t_{\mathrm{c}}-t_i)\right]$, and $h_i=(t_{\mathrm{c}}-t_i)^m\sin\left[\omega\ln(t_{\mathrm{c}}-t_i)\right]$.

We can further reduce the minimisation problem (S5) to be:

$$\hat{t}_{\mathrm{c}}=\arg\min_{t_{\mathrm{c}}}F_2(t_{\mathrm{c}}), \quad \text{(S9)}$$

with

$$F_2(t_{\mathrm{c}})=\min_{m,\omega}F_1(t_{\mathrm{c}},m,\omega)=F_1(t_{\mathrm{c}},\hat{m},\hat{\omega}), \quad \text{(S10)}$$

so that, for a fixed $t_{\mathrm{c}}$, the parameters $\{m,\omega\}$ can be estimated from:

$$\{\hat{m},\hat{\omega}\}=\arg\min_{m,\omega}F_1(t_{\mathrm{c}},m,\omega). \quad \text{(S11)}$$

During the calibration, we employ a filter of $1.5 \le \omega \le 15$, to prevent physically less relevant oscillation scenarios (Filimonov and Sornette, 2013). Here, a relatively low value is adopted for the lower bound to avoid having boundary effect on the model calibration. In addition, log-periodicity may arise due to noise in the data (Huang et al., 2000), where the most probable artificial angular log frequency $\omega_{\text{noise}} = 2\pi f_{\text{noise}}$ can be estimated from:

$$f_{\text{noise}} = \frac{1.5}{\ln(t_c - t_0) - \ln(t_c - t_f)} . \tag{S12}$$

We use the Lagrange regularisation approach (Demos and Sornette, 2019) to detect the optimal start time $t_0$ of the time window for the LPPLS fitting, given a fixed end time $t_f$. More specifically, the following cost function is minimised:

$$\tilde{F}'(t_0) = \tilde{F}(t_0) - \chi N(t_0) , \tag{S13}$$

where $\chi$ is the Lagrange multiplier and $\tilde{F}(t_0)$ is the normalised sum of squared residuals given by:

$$\tilde{F}(t_0) = \frac{F}{N(t_0) - n} , \tag{S14}$$

with $F$ is given by equation (S3) and $n = 7$ is the number of degrees of freedom corresponding to the number of parameters of the LPPLS model. The Lagrange parameter $\chi$ may be approximated via a linear regression of $\tilde{F}(t_0)$ over $t_0$.

**Text S2. Lomb method**

Consider a time series of normalized residual $\varepsilon_i$ sampled at log normalised times $\ln\tau_i$, where $i = 1, 2, \ldots, N$ and $\tau_i = (t_c - t_i)/(t_c - t_0)$. The mean and variance of the signal are respectively given by:

$$\bar{\varepsilon} = \frac{1}{N}\sum_{1}^{N}\varepsilon_i \quad \text{and} \quad \sigma^2 = \frac{1}{N-1}\sum_{1}^{N}\left(\varepsilon_i - \bar{\varepsilon}\right)^2 . \tag{S15}$$

The normalized Lomb periodogram can then be calculated as (Zhou and Sornette, 2002a):

$$P(\omega) = \frac{1}{2\sigma^2}\left\{ \frac{\left[\sum_{1}^{N}\left(\varepsilon_i - \bar{\varepsilon}\right)\cos\omega(\ln\tau_i - \xi)\right]^2}{\sum_{1}^{N}\cos^2\omega(\ln\tau_i - \xi)} + \frac{\left[\sum_{1}^{N}\left(\varepsilon_i - \bar{\varepsilon}\right)\sin\omega(\ln\tau_i - \xi)\right]^2}{\sum_{1}^{N}\sin^2\omega(\ln\tau_i - \xi)} \right\} . \tag{S16}$$

where $\omega = 2\pi f$ is the angular log frequency with $f$ being the log frequency, while $\xi$ can be derived from:

$$\tan(2\omega\xi) = \left(\sum_{1}^{N} \sin 2\omega \ln \tau_i\right) \Bigg/ \left(\sum_{1}^{N} \cos 2\omega \ln \tau_i\right). \tag{S17}$$

In the Lomb periodogram, peaks at specific angular log frequencies suggest the potential presence of log-periodic components, with the highest peak representing the dominant log frequency in the time series. A higher peak indicates greater statistical significance of the corresponding log-periodic component. For data characterised by independently and normally distributed Gaussian noise, the probability that a detected peak surpasses a given height $z$ can be analytically calculated as:

$$p(>z) = 1 - (1 - e^{-z})^{M}, \tag{S18}$$

with $M$ being the effective number of independent frequencies.

On the Lomb periodogram, the maximum peak height $P_{\max}$ can be identified together with the correponding dominant log frequency $f_{\text{Lomb}}$ or angular log frequency $\omega_{\text{Lomb}}$. The false-alarm probability $p_{\text{FA}}$ for the highest peak can be further obtained by substituting $P_{\max}$ for the variable $z$ into equation (S18). We also derive the first-to-second highest peak ratio $\eta$, quantifying the relative significance of the maximum Lomb peak (Zhou and Sornette, 2002a). The signal-to-noise ratio $\gamma$ is also computed as:

$$\gamma = \left(\frac{4P_{\max}}{N - 2P_{\max}}\right)^{1/2}, \tag{S19}$$

which generally holds for various noises (Zhou and Sornette, 2002a, 2002b).

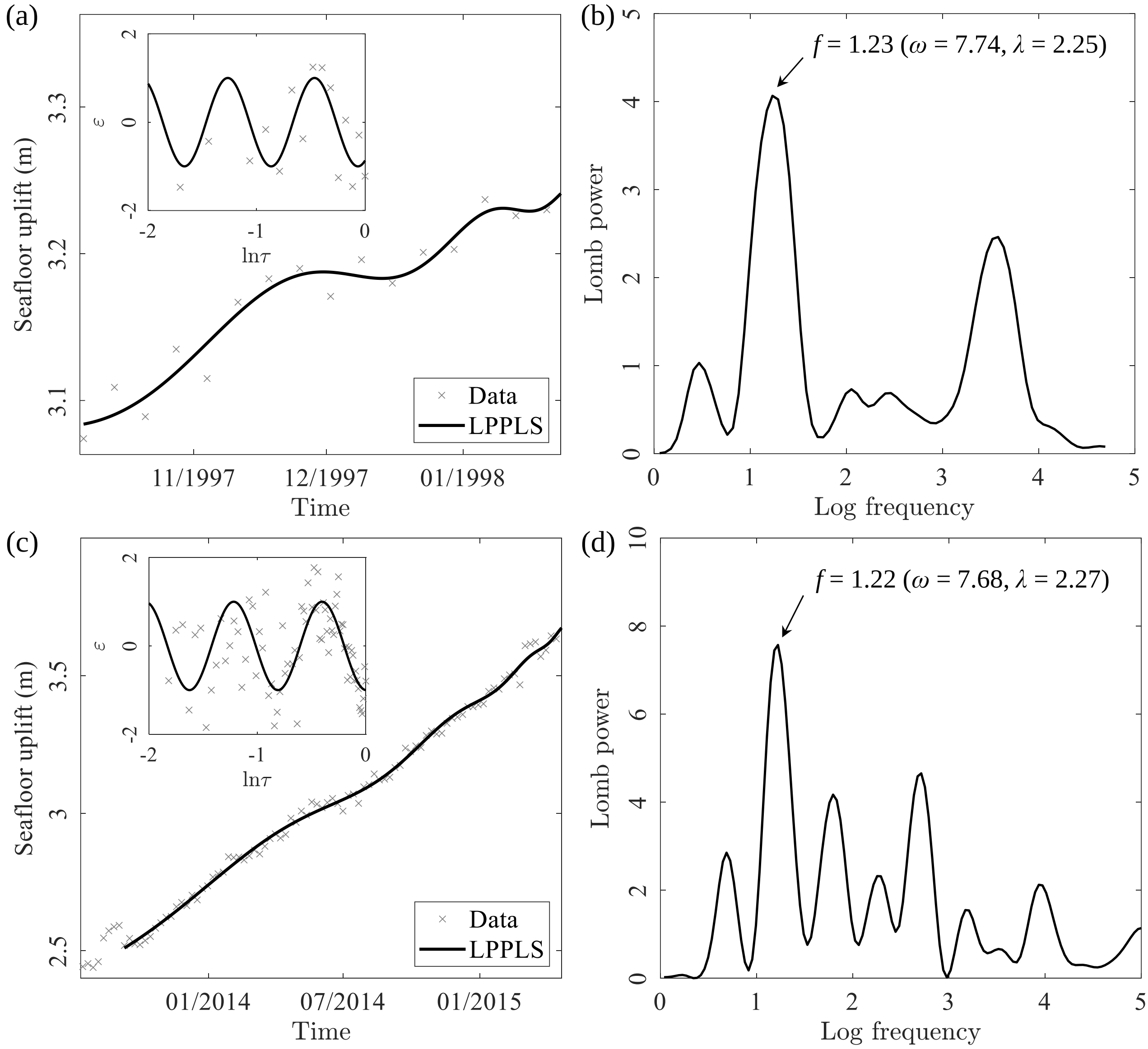

**Fig. S1.** Monitoring data showing the temporal evolution of seafloor uplift (aggregated on a weekly basis) at the central caldera of the Axial Seamount volcano in the Pacific Ocean, prior to its eruptions in (a) January 1998 and (c) April 2015, with (b, d) the corresponding Lomb periodograms for detecting log-periodic oscillatory components in the data. Insets display the temporal evolution of normalised residual $\varepsilon$ as a function of log normalised time $\tau = (t_c - t)/(t_c - t_0)$.

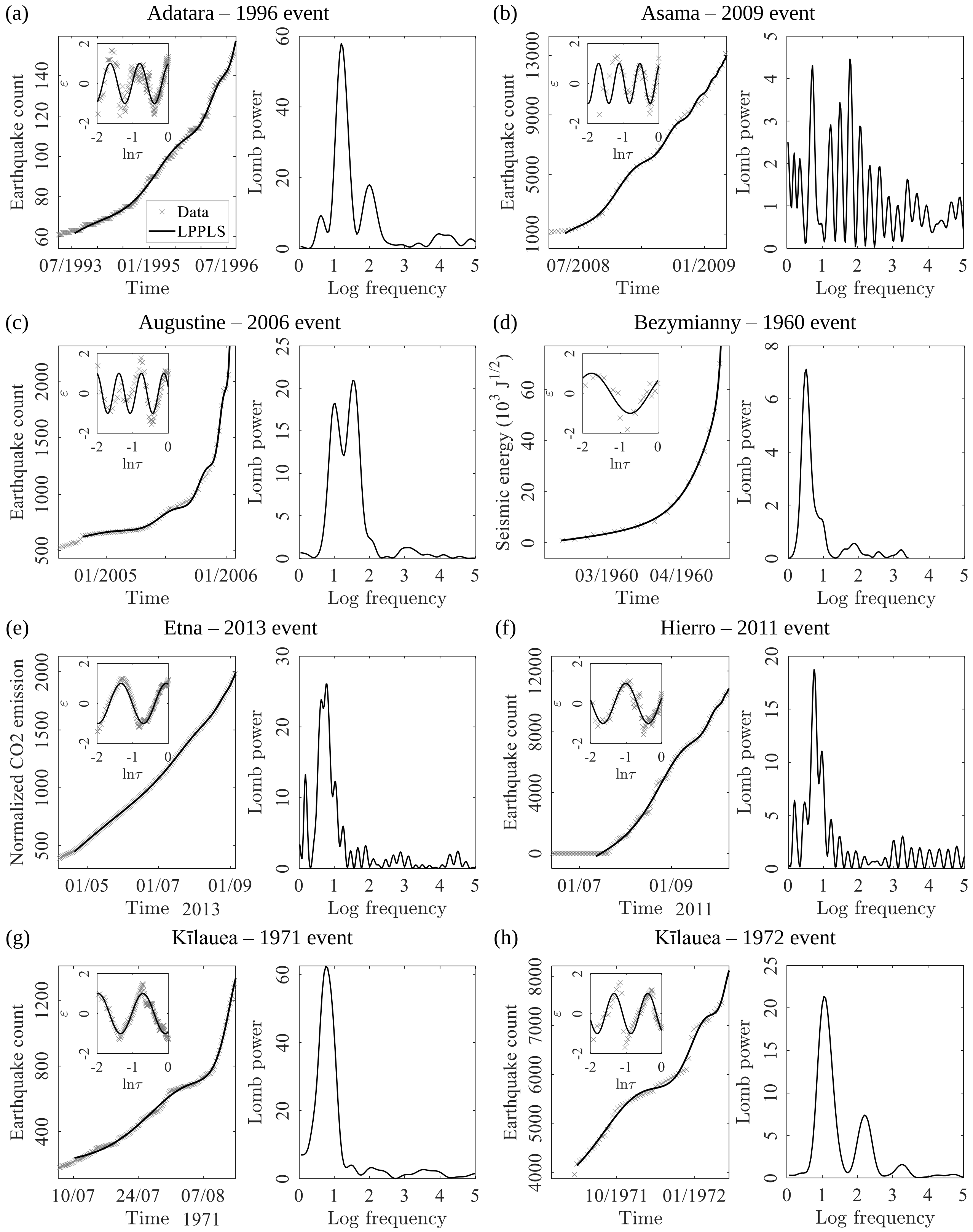

**Fig. S2.** Monitoring data of various volcanoes fitted to the LPPLS model (insets show the variation of normalised residual $\epsilon$ as a function of log normalised time $\tau = (t_c - t)/(t_c - t_0)$) together with Lomb periodograms.

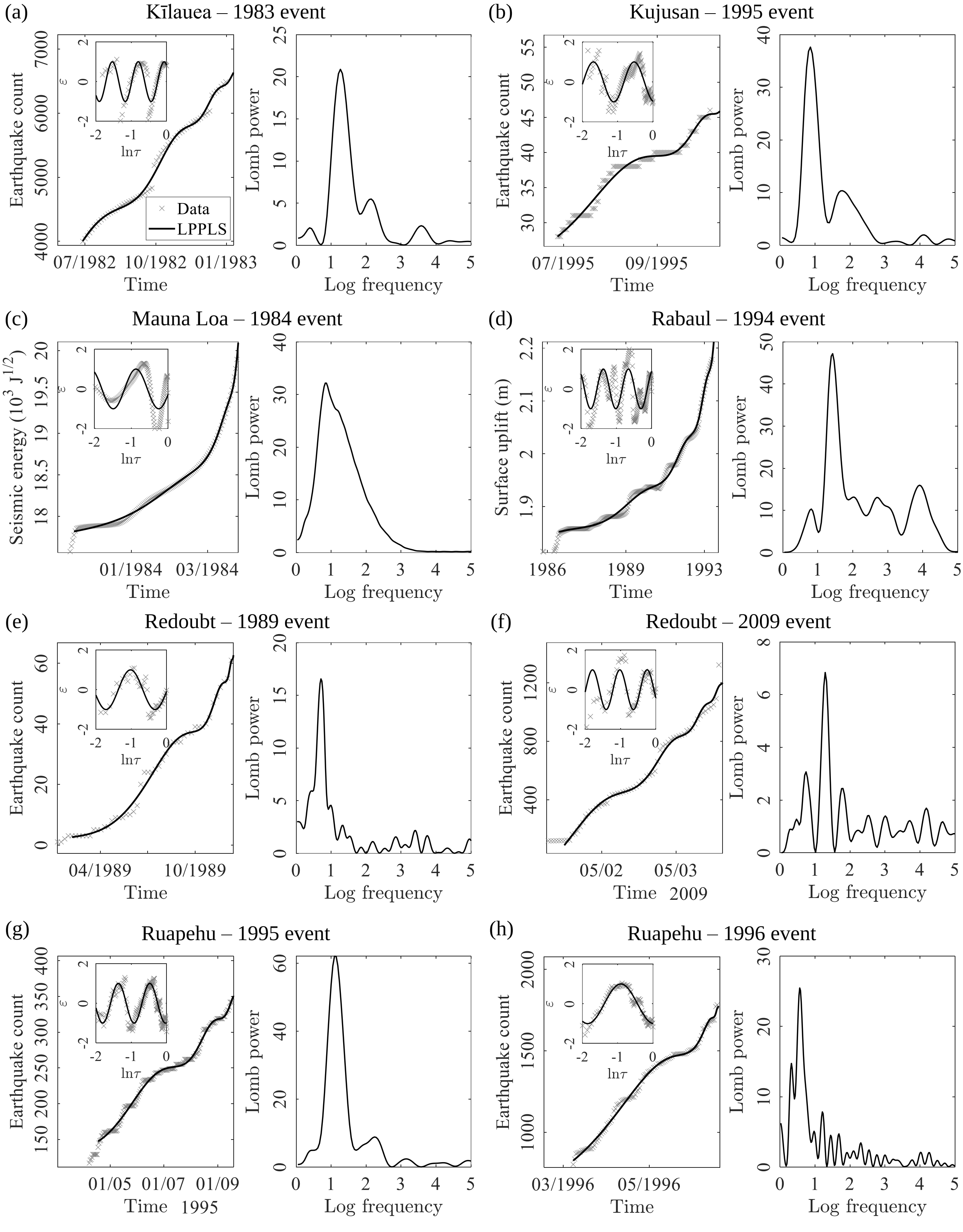

**Fig. S3.** Monitoring data of various volcanoes fitted to the LPPLS model (insets show the variation of normalised residual $\epsilon$ as a function of log normalised time $\tau = (t_c - t)/(t_c - t_0)$) together with Lomb periodograms.

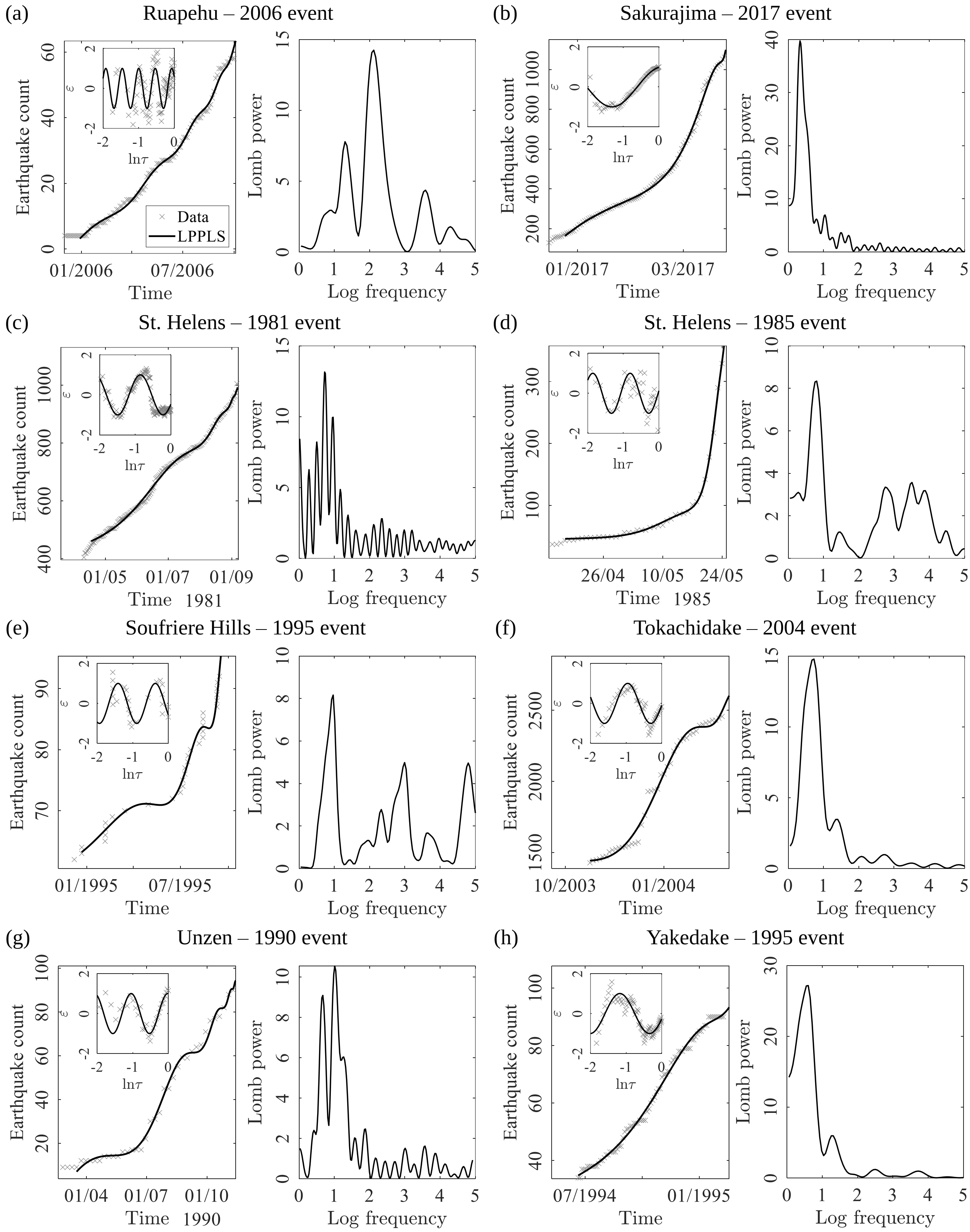

**Fig. S4.** Monitoring data of various volcanoes fitted to the LPPLS model (insets show the variation of normalised residual $\epsilon$ as a function of log normalised time $\tau = (t_c – t)/(t_c – t_0)$) together with Lomb periodograms.

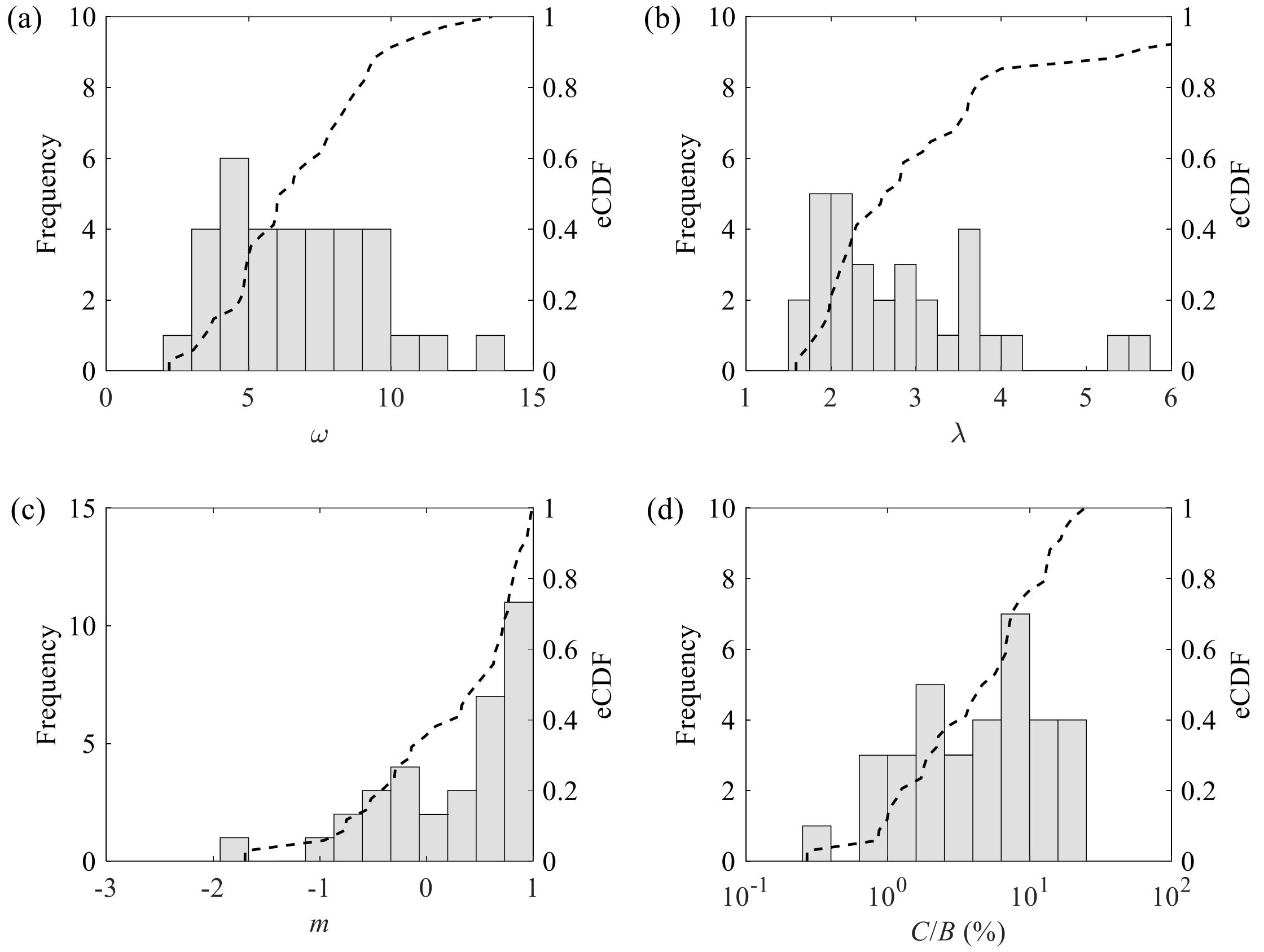

**Fig. S5.** Histograms together with the empirical cumulative distribution function (eCDF) of the LPPLS parameters of 34 volcanic eruption events, including (a) the log-periodic frequency *ω*, (b) the fundamental scaling ratio *λ*, (c) the singularity exponent *m*, and (d) the relative amplitude *C*/*B*.

Table S1. Volcano information (34 events in total).

| Event | Location | Type | Eruption time | Erupted volume ($m^3$) | Monitoring method | Data source | Reference |
|---|---|---|---|---|---|---|---|
| Adatara (1996 event) | Japan | Stratovolcano | 1996-09-01 | $1.1\times10^6$ | JMA seismic network | Original | WOVOdat |
| Asama (2009 event) | Japan | Complex volcano | 2009-02-02 | $\sim10^4$ | JMA seismic network | Original | WOVOdat |
| Augustine (2006 event) | USA | Stratovolcano | 2006-01-11 | $7.3\times10^7$ | USGS seismic network-AVO | Original | WOVOdat |
| Axial Seamount (1998 event) | Pacific Ocean | Shield volcano | 1998-01-25 | $3.1\times10^7$ | Bottom pressure recorders & mobile pressure recorders | Original | (Chadwick et al., 2013) |
| Axial Seamount (2011 event) | Pacific Ocean | Shield volcano | 2011-04-06 | $9.9\times10^7$ | Bottom pressure recorders & mobile pressure recorders | Original | (Chadwick et al., 2012) |
| Axial Seamount (2015 event) | Pacific Ocean | Shield volcano | 2015-04-24 | $1.5\times10^8$ | Bottom pressure recorders & mobile pressure recorders | Original | (Chadwick et al., 2016) |
| Bezymianny (1960 event) | Russia | Stratovolcano | 1960-04-10 | $\sim10^6$ | Unspecified | Digitised | (Voight, 1988) |
| Etna (1989 event) | Italy | Stratovolcano | 1989-09-08 | $\sim10^7$ | ISCSN | Original | WOVOdat |
| Etna (2013 event) | Italy | Stratovolcano | 2013-09-05 | $\sim10^6$ | Geochemical monitoring stations | Original | WOVOdat |
| Hierro (2011 event) | Spain | Submarine shield volcano | 2011-10-10 | $3.3\times10^7$ | El Hierro seismic network | Original | WOVOdat |
| Kilauea (1971 event) | USA | Shield volcano | 1971-08-14 | $9.1\times10^6$ | NCDC-ANSS seismic network | Original | WOVOdat |
| Kilauea (1972 event) | USA | Shield volcano | 1972-02-04 | $1.2\times10^8$ | NCDC-ANSS seismic network | Original | WOVOdat |
| Kilauea (1983 event) | USA | Shield volcano | 1983-01-03 | $4\times10^6$ | NCDC-ANSS seismic network | Original | WOVOdat |
| Kujusan (1995 event) | Japan | Stratovolcano | 1995-10-11 | $2\times10^5$ | JMA seismic network | Original | WOVOdat |
| Mauna Loa (1984 event) | USA | Shield volcano | 1984-03-25 | $\sim10^8$ | Unspecified | Digitised | WOVOdat |
| Merapi (2006 event) | Indonesia | Stratovolcano | 2006-06-06 | $5.3\times10^6$ | Plawangan Observatory | Original | WOVOdat |
| Merapi (2010 event) | Indonesia | Stratovolcano | 2010-10-26 | $\sim10^7$ | Plawangan Observatory | Original | WOVOdat |

Table S1 (continued). Volcano information (34 events in total).

| Event | Location | Type | Eruption time | Erupted volume ($m^3$) | Monitoring method | Data source | Reference |
|---|---|---|---|---|---|---|---|
| Pinatubo (1991 event) | Philippines | Stratovolcano | 1991-06-07 | $5\times10^9$ | Seismic network | Original | WOVOdat |
| Rabaul (1994 event) | Papua New Guinea | Shield volcano | 1994-09-19 | $\sim2\times10^8$ | Geodetic network | Digitised | (Robertson and Kilburn, 2016) |
| Redoubt (1989 event) | USA | Stratovolcano | 1989-12-14 | $\sim10^8$ | ANSS seismic network | Original | WOVOdat |
| Redoubt (2009 event) | USA | Stratovolcano | 2009-03-22 | $\sim10^8$ | ANSS seismic network | Original | WOVOdat |
| Ruapehu (1995 event) | New Zealand | Stratovolcano | 1995-09-18 | $2\times10^5$ | New Zealand seismic network | Original | WOVOdat |
| Ruapehu (1996 event) | New Zealand | Stratovolcano | 1996-06-19 | $\sim10^7$ | New Zealand seismic network | Original | WOVOdat |
| Ruapehu (2006 event) | New Zealand | Stratovolcano | 2006-10-04 | $\sim10^5$ | New Zealand seismic network | Original | WOVOdat |
| Sakurajima (2017 event) | Japan | Stratovolcano | 2017-03-25 | $\sim10^6$ | JMA seismic network | Original | WOVOdat |
| Sierra Negra (2005 event) | Ecuador | Shield volcano | 2005-10-22 | $1.5\times10^8$ | Continuous GPS network | Original | (Geist et al., 2008) |
| Sierra Negra (2018 event) | Ecuador | Shield volcano | 2018-06-26 | $1.4\times10^8$ | Continuous GPS network | Original | (Bell et al., 2021) |
| St. Helens (1981 event) | USA | Stratovolcano | 1981-09-06 | $\sim10^6$ | PNSN | Original | WOVOdat |
| St. Helens (1982 event) | USA | Stratovolcano | 1982-03-19 | $\sim10^6$ | PNSN, tiltmeter | Original | WOVOdat |
| St. Helens (1985 event) | USA | Stratovolcano | 1985-05-25 | $\sim10^6$ | PNSN | Original | WOVOdat |
| Soufriere Hills (1995 event) | UK | Stratovolcano | 1995-11-15 | $7\times10^7$ | ISCSN | Original | WOVOdat |
| Tokachidake (2004 event) | Japan | Stratovolcano | 2004-02-25 | $\sim10^5$ | JMA seismic network | Original | WOVOdat |
| Unzen (1990 event) | Japan | Complex volcano | 1990-11-17 | $2.1\times10^8$ | ISCSN | Original | WOVOdat |
| Yakedake (1995 event) | Japan | Stratovolcano | 1995-02-11 | $6\times10^3$ | JMA seismic network | Original | WOVOdat |

Note: WOVOdat - World Organization of Volcano Observatories (Newhall et al., 2017); PNSN - Pacific Northwest Seismic Network; ISCSN - International Seismological Centre Seismographic Network; JMASN - Japan Meteorological Agency Seismic Network; NCDC - National Climatic Data Center; ANSS - Advanced National Seismic System; GPS - Global Positioning System; USGS - United States Geological Survey; AVO - Alaska Volcano Observatory.

Table S2. Parameters from the LPPLS calibration and Lomb periodogram analysis.

| Event | $m_{LPPLS}$ | $\omega_{LPPLS}$ | $\lambda_{LPPLS}$ | $C/B$ (%) | $f_{Lomb}$ | $\omega_{Lomb}$ | $\lambda_{Lomb}$ | $P_{max}$ | $\eta$ | $p_{FA}$ | $\gamma$ | $f_{noise}$ |
|---|---|---|---|---|---|---|---|---|---|---|---|---|
| Adatara (1996 event) | 0.32 | 7.56 | 2.30 | 1.89 | 1.20 | 7.53 | 2.30 | 57.83 | 3.21 | 0.00 | 2.26 | 0.67 |
| Asama (2009 event) | 0.78 | 10.84 | 1.79 | 3.76 | 1.78 | 11.20 | 1.75 | 4.46 | 1.04 | 0.43 | 0.68 | 0.39 |
| Augustine (2006 event) | -0.96 | 9.95 | 1.88 | 6.77 | 1.53 | 9.60 | 1.92 | 20.94 | 1.15 | 0.00 | 1.95 | 0.64 |
| Axial Seamount (1998 event) | 0.99 | 7.86 | 2.22 | 17.62 | 1.23 | 7.74 | 2.25 | 4.07 | 1.65 | 0.24 | 1.44 | 0.88 |
| Axial Seamount (2011 event) | 0.64 | 9.23 | 1.98 | 6.83 | 1.52 | 9.55 | 1.93 | 46.96 | 1.28 | 0.00 | 0.72 | 0.25 |
| Axial Seamount (2015 event) | 0.81 | 7.71 | 2.26 | 3.56 | 1.22 | 7.68 | 2.27 | 7.57 | 1.63 | 0.04 | 0.66 | 0.57 |
| Bezymianny (1960 event) | -0.55 | 3.06 | 7.79 | 4.64 | 0.51 | 3.21 | 7.07 | 7.12 | 12.61 | 0.02 | 2.22 | 0.51 |
| Etna (1989 event) | -0.3 | 11.84 | 1.70 | 1.02 | 1.87 | 11.77 | 1.71 | 24.00 | 4.56 | 0.00 | 1.23 | 0.72 |
| Etna (2013 event) | 0.76 | 4.73 | 3.77 | 2.22 | 0.79 | 4.96 | 3.55 | 25.48 | 1.08 | 0.00 | 1.07 | 0.30 |
| Hierro (2011 event) | 0.88 | 4.88 | 3.62 | 8.39 | 0.73 | 4.61 | 3.91 | 17.96 | 1.69 | 0.00 | 1.15 | 0.33 |
| Kilauea (1971 event) | -0.016 | 3.61 | 5.70 | 0.27 | 0.56 | 3.53 | 5.94 | 70.00 | 15.26 | 0.00 | 4.18 | 0.50 |
| Kilauea (1972 event) | 0.39 | 6.63 | 2.58 | 4.64 | 1.05 | 6.57 | 2.60 | 21.35 | 2.89 | 0.00 | 2.36 | 0.78 |
| Kilauea (1983 event) | 0.99 | 8.56 | 2.08 | 7.24 | 1.27 | 7.95 | 2.20 | 20.88 | 3.80 | 0.00 | 1.98 | 0.83 |
| Kujusan (1995 event) | 0.77 | 5.44 | 3.17 | 13.42 | 0.87 | 5.45 | 3.17 | 37.63 | 3.63 | 0.00 | 2.37 | 0.81 |
| Mauna Loa (1984 event) | -0.15 | 5.09 | 3.44 | 0.85 | 0.83 | 5.23 | 3.32 | 32.18 | 139.98 | 0.00 | 1.41 | 0.50 |
| Merapi (2006 event) | 0.47 | 4.99 | 3.52 | 1.26 | 0.78 | 4.90 | 3.61 | 21.00 | 4.75 | 0.00 | 3.74 | 0.84 |
| Merapi (2010 event) | -0.75 | 8.36 | 2.12 | 2.27 | 1.42 | 8.91 | 2.02 | 11.05 | 10.87 | 0.00 | 1.85 | 0.85 |
| Pinatubo (1991 event) | 0.088 | 6.56 | 2.61 | 1.15 | 1.06 | 6.64 | 2.58 | 24.73 | 3.64 | 0.00 | 2.52 | 0.45 |
| Rabaul (1994 event) | -0.29 | 8.81 | 2.04 | 1.79 | 1.42 | 8.95 | 2.02 | 47.20 | 2.95 | 0.00 | 1.31 | 0.61 |

Table S2 (continued). Parameters from the LPPLS calibration and Lomb periodogram analysis.

| Event | $m_{LPPLS}$ | $\omega_{LPPLS}$ | $\lambda_{LPPLS}$ | $C/B$ (%) | $f_{Lomb}$ | $\omega_{Lomb}$ | $\lambda_{Lomb}$ | $P_{max}$ | $\eta$ | $p_{FA}$ | $\gamma$ | $f_{noise}$ |
|---|---|---|---|---|---|---|---|---|---|---|---|---|
| Redoubt (1989 event) | 0.55 | 4.53 | 4.00 | 9.86 | 0.71 | 4.46 | 4.10 | 16.56 | 2.79 | 0.00 | 2.36 | 0.39 |
| Redoubt (2009 event) | 0.71 | 8.13 | 2.17 | 5.68 | 1.29 | 8.10 | 2.17 | 6.84 | 2.23 | 0.06 | 0.78 | 0.59 |
| Ruapehu (1995 event) | 0.63 | 7.06 | 2.44 | 7.55 | 1.10 | 6.89 | 2.49 | 62.05 | 7.03 | 0.00 | 2.88 | 0.71 |
| Ruapehu (1996 event) | 0.71 | 3.34 | 6.56 | 16.54 | 0.56 | 3.52 | 5.95 | 25.48 | 1.73 | 0.00 | 1.40 | 0.32 |
| Ruapehu (2006 event) | 0.33 | 13.56 | 1.59 | 0.99 | 2.12 | 13.31 | 1.60 | 14.24 | 1.83 | 0.00 | 0.94 | 0.81 |
| Sakurajima (2017 event) | 0.85 | 2.21 | 17.17 | 24.51 | 0.33 | 2.09 | 20.08 | 39.73 | 5.71 | 0.00 | 3.88 | 0.33 |
| Sierra Negra (2005 event) | -0.52 | 9.10 | 1.99 | 1.72 | 1.51 | 9.51 | 1.94 | 33.10 | 1.70 | 0.00 | 1.61 | 0.81 |
| Sierra Negra (2018 event) | -0.4 | 9.40 | 1.95 | 0.87 | 1.55 | 9.74 | 1.91 | 56.58 | 5.20 | 0.00 | 1.38 | 0.80 |
| St. Helens (1981 event) | 0.72 | 4.91 | 3.60 | 7.06 | 0.73 | 4.56 | 3.97 | 13.47 | 1.33 | 0.00 | 0.68 | 0.30 |
| St. Helens (1982 event, seismic data) | -1.2 | 4.64 | 3.87 | 15.12 | 0.77 | 4.85 | 3.65 | 26.09 | 5.29 | 0.00 | 3.65 | 0.83 |
| St. Helens (1982 event, radial tilt) | -0.31 | 5.83 | 2.94 | 1.61 | 0.99 | 6.23 | 2.74 | 22.24 | 3.11 | 0.00 | 2.32 | 0.83 |
| St. Helens (1982 event, tangential tilt) | -0.77 | 7.15 | 2.41 | 1.88 | 1.16 | 7.26 | 2.37 | 19.70 | 1.98 | 0.00 | 1.87 | 0.83 |
| St. Helens (1985 event) | -1.7 | 5.99 | 2.85 | 13.93 | 0.81 | 5.06 | 3.46 | 8.34 | 2.34 | 0.01 | 1.25 | 0.71 |
| Soufriere Hills (1995 event) | -0.14 | 6.00 | 2.85 | 2.66 | 0.97 | 6.09 | 2.81 | 8.16 | 1.64 | 0.01 | 1.54 | 0.66 |
| Tokachidake (2004 event) | 0.99 | 4.82 | 3.68 | 19.96 | 0.73 | 4.58 | 3.94 | 14.78 | 4.26 | 0.00 | 2.18 | 0.84 |
| Unzen (1990 event) | 0.82 | 6.11 | 2.80 | 12.84 | 1.01 | 6.36 | 2.68 | 10.55 | 1.16 | 0.00 | 1.74 | 0.42 |
| Yakedake (1995 event) | 0.94 | 3.77 | 5.29 | 13.05 | 0.60 | 3.77 | 5.29 | 27.16 | 4.57 | 0.00 | 2.10 | 0.82 |